# Feynman Algorithm Implementation for Comparison with Euler in a Uniform Elastic Two-Layer 2D and 3D Object Dynamic Deformation Framework in OpenGL with GUI

Miao Song
Computer Science and Software Engineering
Concordia University
Montréal, Québec, Canada
m_song@cse.concordia.ca

October 27, 2018

**Abstract**

We implement for comparative purposes the Feynman algorithm within a C++-based framework for two-layer uniform facet elastic object for real-time softbody simulation based on physics modeling methods. To facilitate the comparison, we implement initial timing measurements on the same hardware against that of Euler integrator in the softbody framework by varying different algorithm parameters. Due to a relatively large number of such variations we implement a GLUI-based user-interface to allow for much more finer control over the simulation process at real-time, which was lacking completely in the previous versions of the framework. We show our currents results based on the enhanced framework.

The two-layered elastic object consists of inner and outer elastic mass-spring surfaces and compressible internal pressure. The density of the inner layer can be set differently from the density of the outer layer; the motion of the inner layer can be opposite to the motion of the outer layer. These special features, which cannot be achieved by a single layered object, result in improved imitation of a soft body, such as tissue's liquid non-uniform deformation. The inertial behavior of the elastic object is well illustrated in environments with gravity and collisions with walls, ceiling, and floor. The collision detection is defined by elastic collision penalty method and the motion of the object is guided by the Ordinary Differential Equation computation. Users can interact with the modeled objects, deform them, and observe the response to their action in real-time and we provide an extensible framework and its implementation for comparative studies of different physical-based modeling and integration algorithm implementations.

# Chapter 1

# Introduction

In our real physical world there exist not only rigid bodies but also soft bodies, such as human and animal's soft parts and tissue, and other non-living soft objects, such as cloth, gel, liquid, and gas. Soft body simulation, which is also known as deformable object simulation, is a vast research topic and has a long history in computer graphics. It has been used increasingly nowadays to improve the quality and efficiency in the new generation of computer graphics for character animation, computer games, and surgical training. So far, various elastically deformable models have been developed and used for this purpose. Thus, we introduce the next version of a C++ framework for softbody simulation algorithm to allow for comparison of accuracy and performance of the algorithms to test out softbody simulation concepts outlined in the previous works [8, 9] in general and for Feynman algorithm.

## 1.1 Deformable Object

In engineering mechanics, the term "deformable object" refers to an object whose shape can be changed due to an applied force, such as tensile (pulling), compressive (pushing), bending, or tearing forces. The deformation can be categorized as the following, depending on the types of material and the forces applied (see Figure 1.1):

- Elastic deformation (small deformation) is reversible. The object shape is temporarily deformed when tension is applied and it returns to its original shape when force is removed. An object made of rubber has a large elastic deformation range; silk cloth material has a moderate elastic deformation range; crystal has almost no elastic deformation range.

- Plastic deformation (moderate deformation) is not reversible. The object shape is deformed when tension is applied and its shape is partially returned to its original form when the force is removed. Objects such as silver and gum, which can be stretched at their original length, cannot completely restore their original shapes after deformation.

- Fracture deformation (large deformation) is not reversible, but is different from the plastic deformation. The object is permanently deformed when it is irreversibly bent, torn, or broken apart after the material has reached the end of the elastic deformation ranges. All materials will experience fracture deformation when sufficient force is applied.

## 1.2 Elastic Object

Elastic objects belong to a subset of soft body deformable objects. They are dynamic objects that change shape significantly and keep constant volume in response to collision. They can be bent, stretched, and squeezed. Moreover, they restore their previous shape after deformation. Elastic objects can be divided into two domains:

- Large elastic deformation, such as fluid deformation, which focuses on flows through space. It tracks velocity and material properties at fixed points in space.

- Small elastic deformation, such as tissue deformation, which uses particle systems to identify chunks of matter and track their position, acceleration, and velocity.

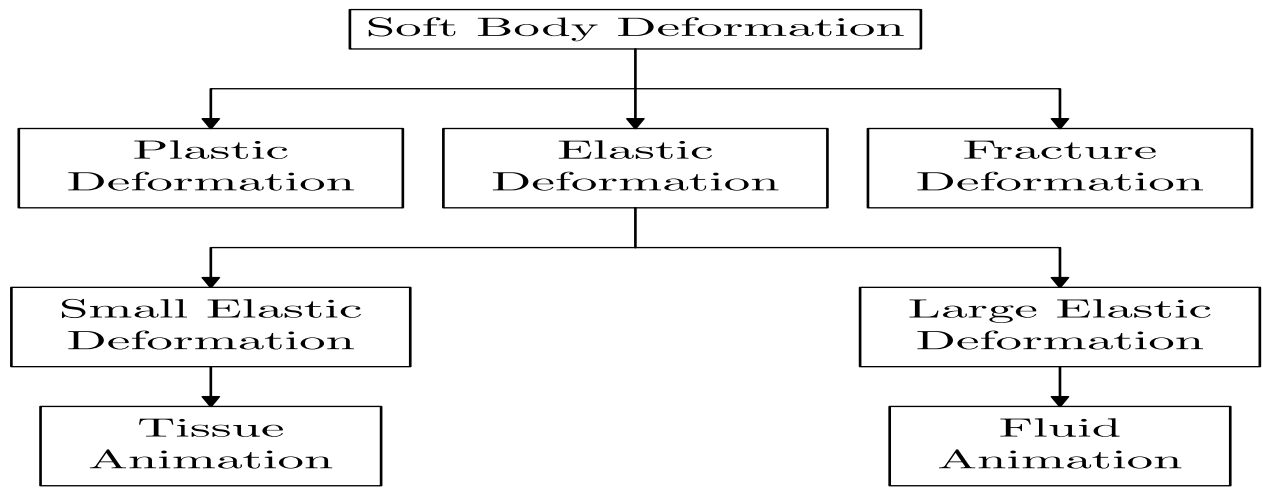

Figure 1.1: Soft Body Deformation

Within this wide research range of soft body simulation, this work has focused on small elastic deformable object simulation, such as tissue animation. Even though there has been many valuable contribution related to this field, there are still many difficulties in accomplishing to realistic and efficient deformable simulation.

## 1.3 Animation Techniques

This section introduces some basic concepts related to the elastic simulation, such as the subject animation method. Animation relies on persistence of vision and refers to a series motion illusions resulting from the display of static images in rapid-shown succession. In traditional animation, squash and stretch are exaggerated for elastic objects. In order to be efficient when working with many of single frame images (or simply frames), inbetweening and cel animation have been introduced by Disney for manual traditional animation [11]. The rate of the animation refers to how many frames are displayed within a given amount of time. If the rate is too low, lower than the brain visual retention, the animation becomes jerky because the brain retains the empty frame from the previous image to the next image. A frame rate, which is the time between two updates of the display, describes the update frequency. In computer games, frames are often discussed in terms of frames per second (fps). The lower bound for smooth animation is between 22 to 30 fps. For many years' research, computer-animation has been developed dramatically to replace the amount of manual traditional animation. The techniques of key-framing, morphing, and motion capture have been widely used [4].

### 1.3.1 Elastic Animation

There are two different methods used for elastic animation modeling, which depends on the predefined simulation or simulation in real time.

**Kinematic modeling**   This technique predefines the positions and velocities of objects. It does not concern what causes movement and how things get where they are in the first place and only deals with the actual movement. For example, given that a ball's initial speed is 10 kilometers per hour on a perfect smooth plane, we can use kinematic method to calculate how far it travels in two hours.

**Dynamic modeling**   This technique, also know as physically based modeling, is the study of masses and forces that cause the kinematic quantities, such as acceleration, velocity, and position, to change as time progresses. For example, when we know the ball's initial speed, we need to know how far it travels after an external force dynamically applied to it.

For elastic object movement, the dynamic methods calculate how the soft body behaves after external force applied dynamically. The animator does not need to specify the exact path of an object compared to using the kinematic modeling method. The system predefines the initial condition of the elastic object, such as position and gravity force. The animation of the object movement is updated each time step based on the acceleration derived from Newton's Laws of motion. The dynamic simulation method is more advanced, easier to achieve the realistic motion than kinematic method. Therefore, we will only represent dynamic simulation of elastic object in this work.

## 1.4   Applications

Elastic modeling has been developed and used in several fields, including geometric modeling, computer vision, computational mathematics, physics engines, bio-mechanics, engineering, character animation, and many other fields [3].

The elastic object for dynamic simulation, which has been widely used, typically has a single-layer elastic surface with different content within. The soft objects can be squashed and stretched according to external and internal forces applied to them. Computation depends on geometric modeling methods and physical equations. However, this method is too inefficient to imitate the behaviors of real human's tissue because human's or animal's soft body does not consist of only one layer with either liquid or air inside.

Soft tissues are multi-composite layers; therefore, one layer elastic object is not sufficient to model the kind of soft body exemplified by human tissue. Moreover, it is difficult to represent the object's inertia caused by the internal material realistically and its liquidity motion based on the various material densities. In this work, we describe a framework for investigation of the accuracy of the two-layer elastic object. The outer layer of the elastic object represents the epidermis and the dermis layer of a real tissue. The objective of this new model is to be visually convincing and to have distinct realism to the animated scene by applying proper physics. The program should be easy to implement, convenient to re-use, and should give best elastic body behavior at the minimum cost rather than model the absolute complex object with the exact accurate physical equations. Users should be able to interact with the soft body in real-time and the collision detection and response must be handled correctly.

# Chapter 2

# Related Work

Research about modeling deformable objects in computer graphics field has been going on for over 40 years and a wide variety techniques have been developed. In this section, we will review the existing geometric approaches for modeling elastic objects. These models are all based on physical laws. From the early elastic model, such as particle model, mass-spring model, finite element model, to recent development such as fluid based model, and pressure model, we briefly introduce their physically-based modeling methods and compare these approaches with their advantages and disadvantages.

**Linear Mass Spring System**  This system has been widely used for modeling elastic objects as shown in Figure 2.1. It is actually derived from the particle model; however, it simplifies the modeling of the inter-particle connection by using flexible springs. Three dimensional systems contain a finite set of masses connected by springs, which are assumed to obey Hooke's Law.

This method was first introduced by Terzopoulos to describe melting effects. Particles, which are connected by springs, have an associated temperature as one of their properties [10]. The stiffness of the spring is dependent on the temperature of the linked two particles. Increased temperature decreases the spring stiffness. When the temperature reaches the melting point, the stiffness becomes zero. The advantages of mass-spring model are that it is easy to construct and display the simulation dynamically. The disadvantages are that such system restricts to only the objects with small elastic deformation with approximation of the true physics, not for the objects that require large elastic deformation, such as fluid. This method also has difficulties to simulate the separation and fusion of a constant volume object. Moreover, the spring stiffness is problematic. If the spring is too weak, for the closed shape model with only simple springs to model the surface will be very easy to collapse. If we try to avoid the collapse, we need to model with spring stiffer, and then we will have difficulty to choose the elasticity because the object is nearly rigid. Another disadvantage is that the mass-spring system has less stability and requires the numerical integrator to take small time steps than FEM model [1].

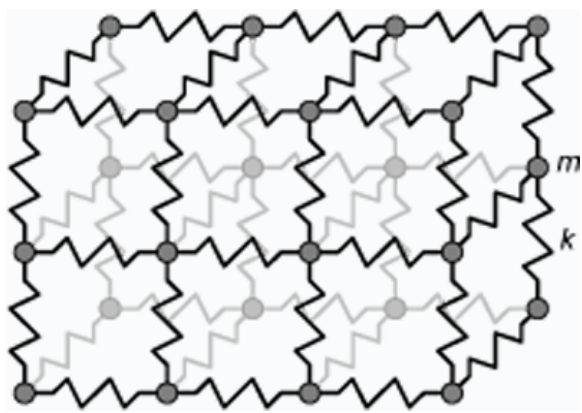

Figure 2.1: Mass-spring Model

**Finite Element Method**  The FEM Model is the most accurate of the physical models used for graphics [3]. It treats deformable object as a continuum, which means the solid bodies with mass and energy distributed all over the object. This continuum model is derived from equations of continuum mechanics. The whole model can be considered as the equilibrium of a general object subjected to external forces. The deformation of the elastic object is a function of these forces and the material property. The object will stop deformation and reach the equilibrium state when the potential energy is minimized. The applied forces must be converted to the associated force vectors and the mass and stiffness are computed by numerically integrating over the object at each time step, so the re-evaluation of the object deformation is necessary and requires heavy pre-processing time [3].

An advantage of FEM model is that it gives more realistic deformation result than mass-spring system because the physics are more accurate. The disadvantage is that the system lacks efficiency. Because the energy equation will be used, the FEM is only efficient for the small deformation of the elastic object, such as application to the plastic material, which has a small deformation range. Alternatively, the object has less control elements needed to be computed, as in cloth deformation. If we need to simulate the human soft body parts or facial animation, the deformation rate is very high. It will be very difficult and sometimes impossible to carry out the integration procedure over the entire body. Therefore, it has been limited to apply in real-time system because of the heavy computational effort (usually it is done off-line). Moreover, the implementation is complicated.

**Fluid-Based Model**  The fluid-based model consists of two components: an elastic surface and a compressible fluid [2]. The surface is represented as a mass-spring system. The fluid is modeled using finite difference approximations to the Navier-Stokes equations of fluid flow. The inner layer is modeled by a particle system, which is similar to real water molecules. Using the numerical methods, the motion of each particle can be computed. In this example, the motion of the each particle is at the center of the basin, and points down to the sink.

The fluid based model uses physically based modeling and it produces realistic fluid animation. It illustrates the behavior of fluid in environments with gravity and collisions with planes. The disadvantage of this model is the heavy computation for both elastic surface and density inside fluid. It also provides a solution to the constant volume problem.

**Pressure Model**  The pressure model was introduced by M. Matyka [6, 5]. It simulates an elastic deformable object with a internal pressure based on the ideal gas law. The object volume is calculated approximately by bounding box, shaped as sphere, cube, or ellipsoid. Another method to determine the object volume is based on Gauss's Theorem.

An advantage of this model is that it gives very convincing effects for elastic properties in real time simulation. The object behaves like a balloon filled only with air. However, the method cannot imitate more interesting effects, such as motion of human tissue. It can not achieve the effect of semi-liquid deformable object because the air pressure density is uniform inside of the object, which is different from liquid with non-uniform density. It is not accurate for describing the inertia of the semi-liquid object.

## 2.1   Feynman vs. Euler Algorithms

In the Euler algorithm, the average velocity and acceleration are replaced by the velocity and acceleration at the beginning of the interval as the equations (1) and (2) where $t$ is the beginning of the time interval, $dt$ is time interval, $v$ is velocity, $a$ is acceleration and $x$ is position:

$$v(t + dt) = v(t) + a(t) \cdot dt \qquad (2.1)$$
$$x(t + dt) = x(t) + v(t) \cdot dt \qquad (2.2)$$

The values at the beginning of the interval are known, and although they are not the best approximation for the average values, they are not bad if the time interval is short enough. The Feynman algorithm approximates the average acceleration and velocity over a time interval by their values at the midpoint (in time). The equations on which the Feynman algorithm are based can be written as equations (3) and (4) using the same notations as for equations (1) and (2).

$$x(t + dt) = x(t) + v\left(t + \frac{dt}{2}\right) \cdot dt \qquad (2.3)$$
$$v\left(t + \frac{dt}{2}\right) = v\left(t - \frac{dt}{2}\right) + a(t) \cdot dt \qquad (2.4)$$

In equations (3) and (4), changes in position are calculated using a velocity value that is half a step ahead in time. Likewise, changes in velocity are calculated using an acceleration which is half a step ahead in time. Position and acceleration are therefore in-phase that is, they are calculated at the same points in time, and velocity is stepped half a step out of phase with both position and acceleration. We can use the Euler and Feynman algorithms to follow the motion of a mass on a spring with assumption that acceleration depends only on time and position.

**Summary** Previous work on deformable object animation uses physically-based methods with local and global deformations applied directly to the geometric models. Based on the survey of the existing elastic models, we conclude their usage as the two types:

- Interactive models are used when speed and low latency are most important and physical accuracy is secondary. Typical examples include mass-spring models and spline surfaces used as deformable models. These can satisfy the character animation with exaggerated unrealistic deformation.

- Accurate models are chosen when physical accuracy is important in order to accomplish the surgical training purpose which requires the accurate tissue modeling. The continuum simulation model, for instance, the most accurate model, FEM, is not ideal for simulation requiring real time interaction and the object undergoing large deformation.

In short, elastic object simulation is a dilemma of demanding accuracy and interactivity where we focus on the latter.

# Chapter 3

# Framework's Design and Implementation Overview

In this section, we present the detailed design in UML of the two-layer elastic object physical based simulation framework and its implementation. The framework's design has centered around common dimensionality (1D, 2D, and 3D) of graphical objects for simulation purposes, physic's based integrators, and the user interactive component all bound by the Model-View-Controller architecture. At present, the Integrator API of the framework is implemented by the well-known Explicit Euler, Feynman, Midpoint, and Runge-Kutta 4 (RK4)-based integrators for their mutual comparison of the run-time and accuracy as of this writing. Thus, we present the supporting components of the framework, their integration and interaction. We plan on releasing our code as open-source implementation either a part of the Concordia University Graphics Library[1] and/or as part of a Maya plug-in.

## 3.1 Elastic Object Simulation System Design

In this section, an overview of the framework's design and the algorithm for the elastic simulation system are given.

### 3.1.1 Domain Analysis-Based Modeling

This elastic object simulation system has been designed and implemented according to the well known architectural pattern shown in Figure 3.1, the Model-View-Controller [12]. This pattern is ideal for real time simulation because it simplifies the dynamic tasks handling by separating data (Model) from user interface (View). Thus, the user's interaction with the software does not impact the data handling; the data can be reorganized without changing the user interface. The communication between the Model and the View is done through another component: Controller. This also closely correlates to the OpenGL state machine, that we use as a core library for our implementation. In our current simulation system, the application has been split into these three separated components:

- Model is an application of object modeling. It stores the geometric modeling methods of the elastic objects and the data of the objects themselves, such as one-dimensional, two-dimensional, and three-dimensional elastic objects and their associated data structures, such as collections of particles, springs, and faces.

- View is the screen presentation to render the Model and a user interface for dynamical simulation. The View in our system is the GLUT window which displays the elastic object and allows the user to use mouse and keyboard to interact with the elastic object. It is to be extended with a richer GUI to allow expert users to exercise more control of the simulated environment.

- Controller handles the processes and responds from the user interaction and invokes the changes to the Model. When the user interacts with the elastic object through the GLUT window by dragging it with mouse, the controller handles the new dragging force from the user interface, integrates the new force to find out the change of the acceleration and velocity, and where the object should move to in next display update. This is done through the series of registered GLUT callback functions that process the input from the user.

---

[1]http://users.encs.concordia.ca/∼grogono/Graphics/cugl.html.

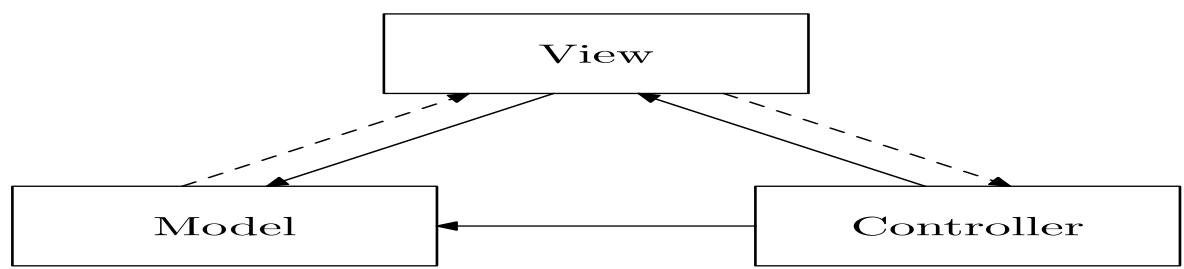

Figure 3.1: Model-View-Controller

## 3.2 Elastic Object Framework and Simulation System Implementation

The system is implemented using OpenGL and the C++ programming language with object oriented programming paradigm. Figure 3.2 presents structure of the software based on the classes.

- The three data structures, such as "Particle", "Spring", and "Face" compose an elastic object.

- The elastic object types can be varied by the dimensionality: one-, two-, or three-dimensional.

- The types of integrators are also varied by their complexities, such as Euler, Feynman, Midpoint, and Runge-Kutta 4.

- An "Object" instance contains an instance of an "Integrator". The relationship between them is aggregation rather than a common composition because when the elastic object is destroyed, the integrator object is not necessary destroyed. The "Object" has an aggregation of the "Integrator" by containing only a reference or pointer to the "Integrator".

- The classes "Object", "ViewSpace", and "Integrator" are associated to each other based the Model-View-Controller model.

We now take a closer look at each model and the related classes with their parameters and member functions.

### 3.2.1 Design and Implementation of Data Types

The basic data structure is the object vector, which defines the the scalar value with direction. For the second basic data structure, particle, whose properties, such as position, velocity are made up of the object vector. The next higher data structure is spring, which is defined by two particle objects. "Face", which is the highest data structure in this simulation system, is composed of three connected springs.

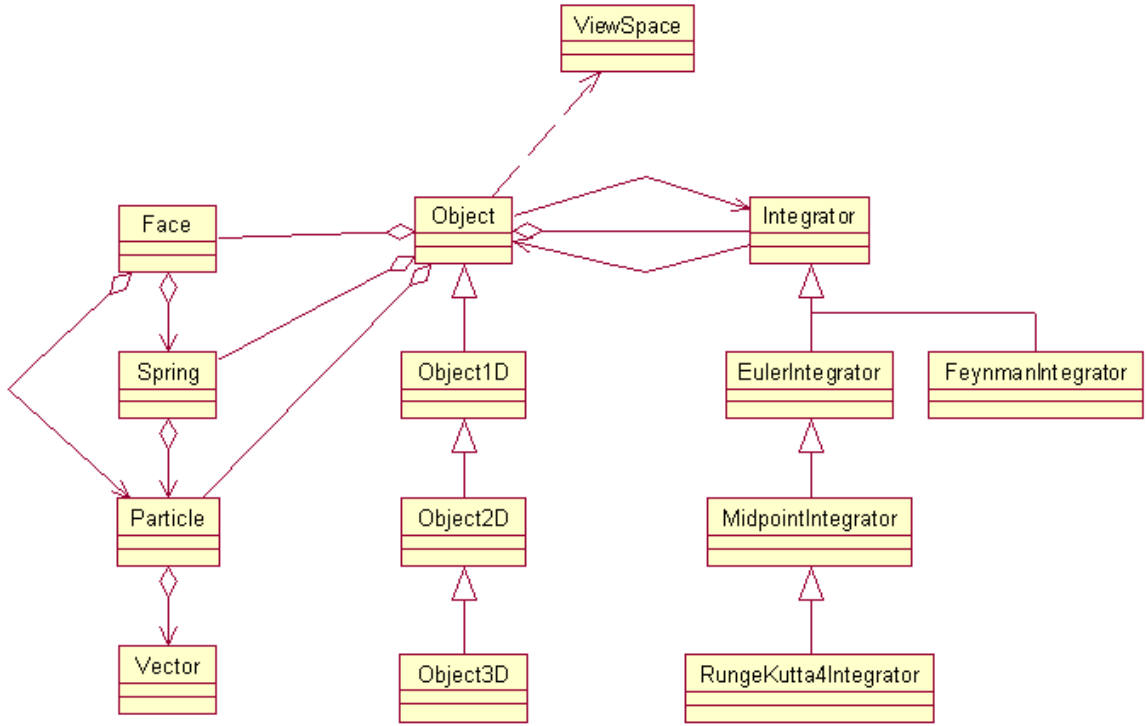

Figure 3.2: Class Diagram of the Main Framework's Components

**Particle**  The particle class has variables for mass $mass$, position $\mathbf{r}$, velocity $\mathbf{v}$, derivative of position $\mathbf{dr}$, derivative of velocity $\mathbf{dv}$, and force vector $\mathbf{f}$. The constructor sets up its properties with default values.

**Spring**  The spring class describes the various different types of springs used to construct the object, such as structural, radius, shear-left, and shear-right springs. The default spring type is structural. $*sp1$ is the head of the spring and points to a particle; $*sp2$ is the tail of the spring and points to a particle. $restLen$ is the spring length when it is in the resting state. $ks$ is Hooke's spring constant and $kd$ is the spring damping factor. The spring normal vector will be calculated and needed in pressure force calculation.

**Face**  The face class contains variables $*fp1$, $*fp2$, and $*fp3$ pointing to the first, the second, and the third particles as three of its vertices. It also contains $*fs1$, $*fs2$, and $*fs3$ point to the first, second, and third spring as three of its edges. There are two face constructors. The first constructor stores the information of three vertices that point to three particles. It represents faces on two-dimensional objects. The faces will only be needed at the display process. The second constructor accepts three vertices on each face that point to the three particles, and constructs a spring and stores the spring information into the spring vector. This constructor is called by three-dimensional uniform modeling method. The index of face is the key data structure for subdivision method in subroutine. The constructor initializes the three springs based on the three particles. First spring contains particle $p1$ and $p2$; the second spring contains particle $p2$ and $p3$; the third spring contains particle $p3$ and $p1$.

Special care is taken not to duplicate existing springs (which would result in incorrect behaviour of the model); therefore, we only allow the new and non-existing springs to be saved in the spring vector. If the first spring already exists with particles $p1$ and $p2$, the new spring $fs1$ will point to the existing spring. Same method is applied on the second spring $fs2$ and third spring $fs3$. Otherwise, the new spring will be pushed and saved into the spring vector.

### 3.2.2 The Model Component

The class "Object" is the base class for elastic object of any supported dimensionality. It contains the most common data structures and properties of an elastic object. The geometric complexity increases according to the dimensions. The "Object1D" inherits from the parent class "Object", "Object2D" inherits from "Object1D", and "Object3D" inherits from "Object2D". This type of inheritance hierarchy is in place because when each dimensionality is added, the new object type depends on some of the previous implementation and the new properties and behaviours that come with each additional dimension. For example, 1D object has a notion of structural springs varying in a single dimension; 2D takes the notion of structural springs and augments it with radius and shear springs as well as the notion of pressure inside an enclosed object; 3D extends 2D by adding the notion of face subdivision and volume making object more dynamic in terms of run-time number of vertices (to make it more or less smooth depending on the trade off between quality and performance). All objects share the same $Update()/Draw()$ API, which is used by the OpenGL state machine to update all the vertices of an object in the Model and reflect the changes in the View by drawing the deformations in real-time.

**Object** The object class represents an elastic object containing a particle object, a spring object, a face object, and an integrator object. The data structures vary from inner to outer layers, for example, the pointers to the particles on the inner layer and on the outer layer of the object are saved in different data vectors. $SetObject()$ constructs the geometric shape of the elastic object, which, in turn, constructs the particles $SetParticles()$ and connects the particles by the structural springs via the $Add\_Structural\_Spring()$ method call. The enumerated type $dimensionality$ has one of the values ($DIM1D, DIM2D, DIM3D$) to determine the object's dimensionality type: 1D, 2D, or 3D; the enumerated type $integrator\_type$ determines which type of integrator the simulation system uses, Euler, Midpoint, or Runge Kutta Fourth Order integrator. Such design allows extension to add new integrators and select existing integrators at run-time. The variable $closest_i$ is the closest point on the outer layer to mouse position and $FindClosestPoint()$ is the function to find such a particle (used in dragging force application when dragging the object across the simulation window). The function $Update()$ modifies the simulated object's state (either each time point when idle or application of the drag force by the user), and determines the object's overall forces, velocity, position in the next time step. $Draw()$ visualizes the object after each update and is typically invoked from the OpenGL's $display()$ callback.

In the main simulation, the $Idle()$ function shown in Figure **??**, elastic objects update at every time step $DT$ to tell the the system how the objects behave and the change for their velocity and position. There are four parameters for $Update()$ as shown in Figure **??**, the time step $deltaT$, if there exists user interaction $drag = 0$ by default, the mouse position on $x$ and $y$ axises (for dragging upon mouse release) is at 0 by default. The general algorithm of the $Update()$ presented, illustrates that the most of the actual modifications are based on the dynamically selected integrator and the dimensionality of the simulation object being integrated. If in the feature a new integrator is added, this function has to be updated to account for it in the framework.

**2D Object** The Object2D class represents an two-dimensional object that contains inner and outer layers. The type of particles is $inner\_points$ and $outer\_points$. The spring type is structural $inner\_springs$ and $outer\_springs$; moreover, there are another three new types of springs, $radius\_springs$, $shear\_springs\_left$, and $shear\_springs\_right$. The function $Add\_Structural\_Spring()$ models the shape of the inner circle by connecting $inner\_springs$ and the outer circle by connecting the $outer\_springs$ separately. $Add\_Radius\_Spring()$ adds the radius springs with the inner point $i$ and outer point $i$. $Add\_Shear\_Spring()$ adds the left shear springs with inner point $i$ and outer point $i + 1$ and the right shear springs with inner point $i + 1$ and outer point $i$. The variable $pressure$, which is an additional inner force compared to "Object1D", is at each spring along its normal.

**3D Object** This class represents a three-dimensional object that uses methods similar to the 2D object but extends the variables into the $z$ axis. Two methods are introduced to create a three-dimensional object, *nonunitsphere()* and *SetObject()*, which uses iteration to define an uniform sphere. The base shape for subdivision a sphere is defined in *Octahedron()* and *Iteration()* computes the coordinates of the newly generated particles and springs based on the level of detail, the variable *Iterations*.

### 3.2.3 The Controller

The Controller's architecture connects the Model and the View via the integrators that the data has to pass through when the user interacts with the softbody on the screen as well as the softbody's collision response is taking place. The part of the Contorller mechanism is invoked through the OpenGL callbacks attached to the mouse and keyboard causing the change of forces applied to the softbody and, as a result, re-integration of all the force components. The types of integrators are varied by their complexities, such as Euler, Midpoint, and Runge-Kutta. The common attributes and methods are defined in the parent class "Integrator". The subclasses "EulerIntegrator", "MidpointIntegrator", "RungeKuttaIntegrator", and the new "FeynmanIntegrator" inherit the super classes based on the complexity. In the present form, the Euler integrator is a basic building block for other integrators which provides the first step of computation of $k_1$ in $k1()$. Midpoint integrator uses Euler's $k1()$ implementation and provides the 2nd step, $k_2$ implemented in $k2()$. Finally, the RK4 integrator adds the last two refinement steps $k_3$ (function $k3()$) and $k_4$ (function $k4()$) in addition to what Euler and midpoint have provided. Thus, RK4 implementation depends on the midpoint which, in turn, depends on the Euler integrator with different parameters. Such a dependency need to be the case for all integrators, we can accept any integrator plug-in so long it implements the "Integrator" API.

Class Integrator has an *integrate()* method, shown inFigure **??**, which is called from $Object::Update()$, and an *AccumulateForces()* method. Both of these methods play a vital role in the integrator framework in this work. They illustrate the general algorithm of integration applied to the Model's data: first, the effect of all the forces is accumulated (which includes external forces, such as gravity and drag, as well as forces induced by springs and pressure); then, the integrator-specific derivation is performed to each particle of an object. In the general Integrator the *Derivatives()* method is pure virtual as is left to be mandatorily overridden by the EulerIntegrator, MidpointIntegrator, and RungeKutta4Integrator concrete implementations. It is important to note that the reverse forces are also accounted at the collision detection at the end of each *Derivatives()* implementation.

*ExternalForces()* checks for the existence of the mouse drag force (from the user) as well as gravity and sums them up. *SpringForces()* accumulates contributions for all spring types.

### 3.2.4 Simulation Loop Sequence

The UML sequence diagram in Figure 3.3 describes the control-flow of the simulation sequence and logic of the elastic object simulation system based upon our framework. The following sequence of steps describes all of the possible states of the elastic object as events occur in greater detail. There we track the different states how the physical simulation loop works, such as display of the objects, accumulation of forces, integration of forces, and so on. In other words, this is the main algorithm of the entire simulation system.

1. "ViewSpace" initializes the virtual world and provides the user an interactive environment. It provides the interface to allow user to drag the object, or choose the parameters.

2. *SetObject()* function creates an elastic object based on the interface variable set from 1.

3. *SetParticles()* function sets up the particles' position and their other initial properties, such as mass and velocity.

4. *AddSprings()* function connects particles with springs according to their index.

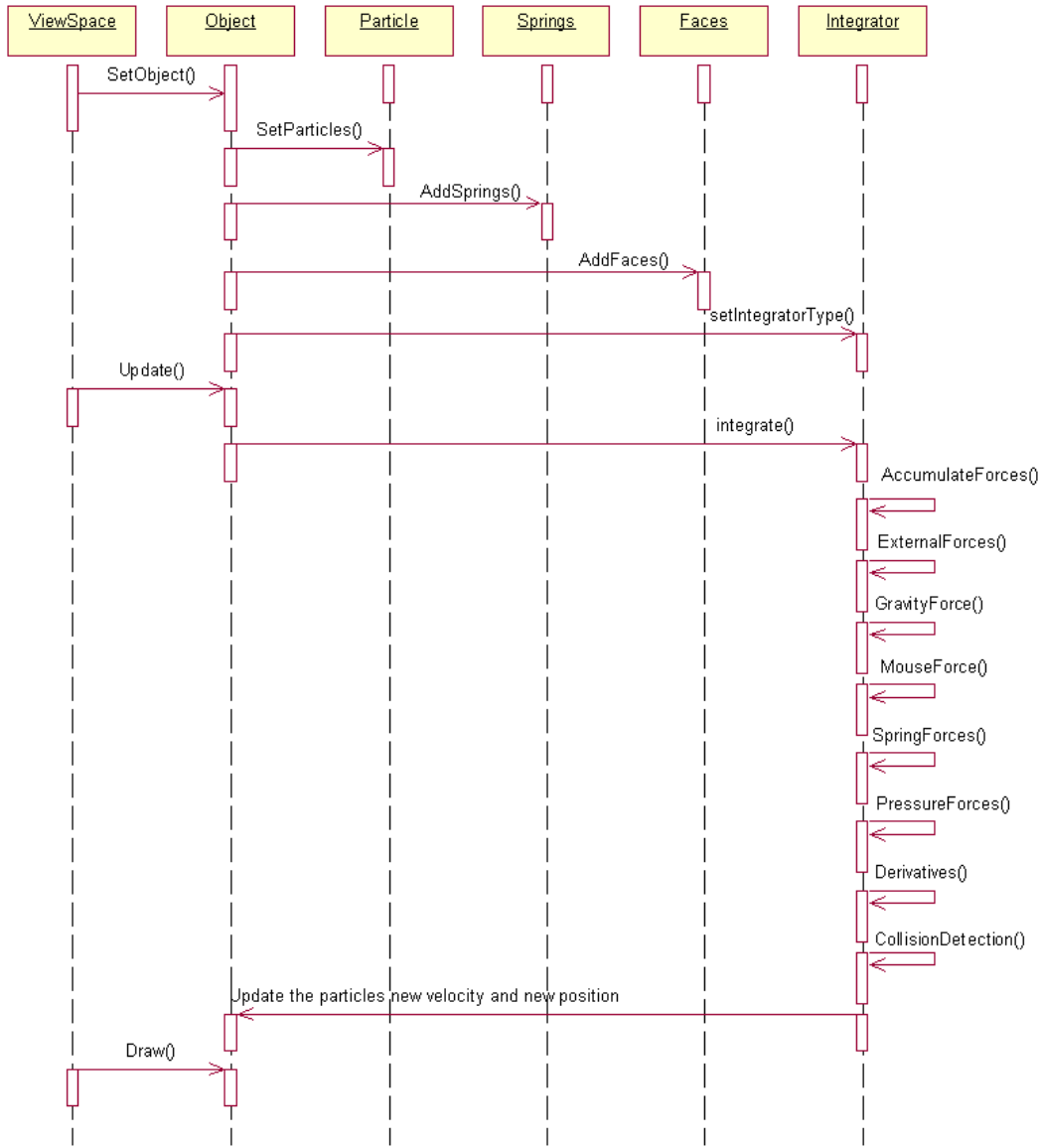

Figure 3.3: Simulation Loop Sequence Diagram

5. *AddFaces()* connects the springs with faces based on proper index. This step will be ignored if the object is one-dimensional.

6. *SetIntegratorType()* function tells the Controller which integrator users select through the interface.

7. *Update()* updates the integrator's time step.

8. *Integrate()* contains two functions, *AccumulateForces()* and *Derivatives()*. It is based on all the object geometric information modeled and all the forces information accumulated, to integrate over the time step to get new object position and orientation.

9. *AccumulateForces*() state is to sum up the forces accumulated on each particle.

10. *GravityForce*() is to accumulate gravity force based on the particles' masses.

11. *MouseForce*() is the external force from the interface when user interacts with the object. It will be added or subtracted from the particles depends on the force's direction.

12. *SpringForce*() is to accumulate internal force of the particles connected by springs.

13. *PressureForce*() is to accumulate the internal pressure acted on the particles. For one-dimensional object, this state is omitted.

14. *Derivatives*() does the real derivative computation of acceleration and velocity in order to get new velocity and position of elastic objects based on the integrator type defined by users.

15. *CollisionForce*() is to check if the object is out of boundaries after the integration state. If the new position is outside of the boundary, then it will be corrected and reset on the edge of the boundary. Moreover, the new collision force will be added to the object.

16. *Draw*() displays the object with new position, velocity, and deformed shape.

### 3.2.5 GUI

The new initial GLUI-based user interface is another new addition in this work. It allows the user to conveniently alter most of the simulation parameters at real-time and take measurements much easer. The previous versions of the framework required recompilation each time when parameters change. See the Experimental Results chapter and the actual code and a running demo for the example of the GUI. In a nutshell, we allow changing particle mass, Hooke spring and damping forces, mouse drag forces, level-of-detail (LOD) of the objects in terms of number of particles (and springs as a result), altering pressure, integration time step, and selecting one of the implemented integrators at run-time. GUI also now allows the 1D, 2D, and 3D objects to appear at the same time on the same simulation scene and allows dragging all three of them by a mouse.

# Chapter 4

# Experimental Results

In this section, the one-, two-, and three-dimensional objects are illustrated at different animation sequences, with different simulation parameters, and by simulation with different numerical integration methods.

## 4.1 Animation Sequence

The screenshots for this section appear at the end of the work. They present the animation sequence of the two-dimensional, and three-dimensional objects when they are at the initial state, colliding with floor, bouncing back from the floor, responding to user's external dragging, and at the resting state.

### 4.1.1 1D

One-dimensional object is nothing but a single spring connecting two particles. Normally one end of it should be fixed onto a surface, but we allow a freedom of movement to drag it around and see how it bounces off the walls, stretches, squeezes, and so on – this is a baseline for our modeling.

### 4.1.2 2D

Figures 5.1(a) through 5.1(f) show how a two-dimensional object moves in a three-dimensional environment. This two-layer object consists of 10 particles and 10 structural springs on both inner and outer circles. Moreover, it contains 10 radius springs, 10 shear left springs, and 10 shear right springs between the inner and outer layers. If a two-dimensional object with only one layer, or the object has no pressure force within, the spring's stiffness has to be a larger value than without, then the object will not collapse. However, as shown in Figure 5.1(b), if the spring stiffness is small enough, the object does not collapse, neither overlap with the layers because of the stability of the two-layer structure.

### 4.1.3 3D

The simulation as shown in Figures 5.2(a) through 5.2(f) is how a three-dimensional uniform facet object moves in a three-dimensional environment. This two-layer object, which is generated by subdividing an octahedron once, consists of 12 particles, 36 structural springs, and 32 faces, on both inner and outer spheres. Moreover, the object also contains 36 radius springs, 36 shear left springs, and 36 shear right springs between the inner and outer layers. Just like in two dimensions, the two-layer structure gives the three-dimensional sphere more stability.

## 4.2 Simulation Parameters

The parameters in the simulation such as mass, spring stiffness, and friction (damping) can be changed. One can drag the object mass with a mouse to change its position. Effects of different simulation parameters are discussed.

### 4.2.1 Summary of the GUI-Adjustable Parameters

The parameters that influence the behavior of the simulated environment are summarized below, with their default values. Most initial and default values were based on the 2D case from [6]; otherwise, the values are empirical and are partially dependent on the hardware the simulation is executing on.

- KS = 800 where KS is structural spring stiffness constant. The larger this value is, the less elastic the object is and it is more resistant to the inner pressure and deformation. The lesser this value is the more object is deformable and a subject to break up if the inner pressure force is high.

- KD = 15 where KD is structural spring damping constant, opposite to the spring retraction force. It denotes how fast the object is to resist its motion.

- RKS = 700 where RKS is radius and shear spring stiffness constant, similar to KS, but for radius and shear springs as opposed to the structural springs.

- RKD = 50 where RKD is radius and shear spring damping constant, similar to KD, but for radius and shear springs.

- MKS = 150 where MKS is the spring stiffness constant of the spring connected with the mouse and the approximate nearest particle on the object. This constitutes the elasticity of the "drag" spring connected to the mouse: the lesser the value is, the more elastic it is, and the harder it is to drag the object as a result.

- MKD = 25 where MKD is the damping constant of the spring connect with the mouse and the approximate nearest point on the object.

- PRESSURE = 20 where PRESSURE is gas constant used in the ideal gas equation mentioned earlier to determine the pressure force inside the enclosed object. If this constant is too high, and the combined spring stiffness for all the spring types is low enough, the object can "blow up".

- MASS = 1 where MASS is the mass for each particle. The object can be made heavier or lighter if this value is larger or smaller respectively, in order to experiment with the gravity effects. Naturally, the heavier objects will be more difficult to drag upwards in the simulation environment. Conversely, the smaller-mass object can be dragged around with less effort given the rest of the parameters remain constant.

### 4.2.2 Stability vs. Time Step

First, the figures in this section (5.3(a), 5.3(b), and 5.3(c)) show the stability of the three integrators. We consider the integration time step parameter in these scenarios only, assuming all the other parameters (discussed later) are not change for the described simulations. As shown in those figures, when the time step is small, such as $DT = 0.003$[1], three of the integrators behave well and the object does not "blow up". However, when one increases the time step by a factor of 10 to $DT = 0.03$, the midpoint (see Figure 5.4(b)) and RK4 (see Figure 5.4(c)) integrators are still stable and the object integrated with Euler integrator "blows up" as in Figure 5.4(a). The Feynman-based integrator, which is technically on the same level as Euler is capable om holding the object together a little longer due to the half-step approach. Furthermore, when the time step is increased 10-fold more to $DT = 0.3$, only the object integrated with RK4 (see Figure 5.5(c)) is stable and another two objects integrated with Euler (Figure 5.5(a)), Feynman, and Midpoint (Figure 5.5(b)) methods "blow up".

---

[1]This is an empirical value; dependent on the performance of the hardware.

### 4.2.3 Efficiency and Accuracy

The more computational effort is required, the less efficient algorithm is. Likewise, the more accurate algorithm is, the more computation effort it requires, the less efficient it is. Thus, in our simulation system the most efficient and least accurate integration method is Euler's, followed by Midpoint (about twice as more accurate and slower), followed by RK4 (four times slower than Euler's and the most accurate of the three). This can be illustrated in Figures 5.3(a), 5.3(b), and 5.3(c) running concurrently with the same time step of 0.003, where one can see the simulation with Euler's method reaches the floor fastest and RK4 slowest. Of course, the efficiency of the simulation and the accuracy of the shape and movement depends on the amount of particles (and as a result, all kinds of springs) in the object.

### 4.2.4 GUI

Screenshots showing the new GUI that allows manipulation object parameter at real-time are in Figure 5.7, Figure 5.8, Figure 5.9, and Figure 5.10 respectively.

## 4.3 Computational Errors

This section briefly summarizes the error accumulated in the application of the described algorithms and their effects.

### 4.3.1 Collision Detection

We have applied the Penalty Method in our simulation system. This simple but inaccurate algorithm causes the object to "stick" on the collision surface when dragging the object at the same time and it may become difficult to drag the object away for a period of time.

### 4.3.2 Subdivision Method

The spherical shape is not perfectly round because the number of springs associated to each particle is not uniform. If one wants more quality subdivision has to be done in more than one subdivision operation, but the simulation may rapidly become very slow as the number of particles grow requiring a much greater computational effort, which is suitable only for the high-end hardware if one wishes to do it in real-time. In Figure 5.6 is an example of the two iterations of the subdivision.

## Chapter 5

# Conclusions and Future Work

This section describes our contribution based on the existing elastic model and analyzes the possible development and related work in the future.

## 5.1 Contribution

The new model, two-layer elastic object with uniform-surfaces is a simple, efficient approach to imitate the liquid effects of elastic object, such as human's tissue and soft body. Since the modeling and structure of the tissue kind elastic object is closer to real tissue than an one layer object, the level of realism has been increased. The modeling method and the density setting provides significant improvements on the conflicts of accuracy and interactivity on previous models. The realism of the results, such as liquid motion and inertia effects are also enhanced. We are able to demonstrate that even with our initial framework design and its implementation. The framework relies on several background concepts such as Procedural Modeling, Density and Inertia properties of two-layered design, the Stability of our Models, and the re-usable object-oriented software implementation.

**Procedural Modeling**  We have applied the procedural modeling method with particle system to model elastic objects. From simple one-dimensional to most complicated three-dimensional object, we introduced the modeling method for different dimensional objects and related physics knowledge gradually. In the elastic object simulation system, each particle has its local coordinate which is easy to be computed at every time step. Moreover, this modeling method can efficiently control the level of detail as required by graphics artists and computer hardware available. This modeling method also most approximately approaches the ideal equal faces; therefore, the edges (springs) on the faces and the forces on each particle are approximately to be equal at initial state in order to minimize the computation error caused by the object geometry.

**Density**  The density is defined only for each particle on the elastic surface and the internal density is represented by air pressure physics equation. The weights of particles on inner and outer layer can be set differently. For example, a balloon half filled with liquid, the bottom is heavier than the top part because the density is at the bottom is liquid and top part is air. The weights on inner layer can be set much heavier than outer layer. This special feature gives us flexibilities to imitate different material effects with such simple model.

**Inertia**  Inertia effect is a unique effect in two layer-elastic simulation system, which can not be achieved with one-layer object. The inner layer and the outer layer have the opposite internal force drive them along axis $x$. Since the two layers are connected by springs, the inner particles and outer particles have an extra force applied on them, interactive force between inner and outer particles. And their movement, position, and acceleration will be computed according to the contribution of this extra interactive force. This interactive force does not exist in a single layer object. The outer and inner particles will fall with the object based on their gravity and springs force. Here, the inertia for inner particle and outer particle are dependent not only on the force from their own motion, the force from the neighbors on the same layer, but also from the interaction on the other layer. This simulation system is more accurate to describe the inertia property happened in the liquid object.

**Stability** The two layered system is stable. Even without the internal pressure force, the shape will not collapse because the two layers are connected by different types of springs. The simulation system works well even with the very inaccurate Euler integrator at large time step, which will result shape collapse or blow up on a one-layer object with the same set of values. We have also implemented the higher level integrators, such as Midpoint and Runga Kutta 4.

**Re-usability** The design of this simulation system is based on well-known software design pattern. It decomposes the novel concepts into concrete small components. The functions and classes are easy to be plugged and adapted into other program. This elastic simulation model simplifies the physical modeling method with a group of masses and springs. Also, the simulation is computed in real time based on the numerical integration of the physical laws of dynamics.

## 5.2  Future Work

- Character Animation. The functionality development of elastic simulation modeling for 3D software design and implementation has emerged as a new challenge in computer graphics. One of the existing software with the elastic modeling functionality is Maya, which provides shape deformation, especially facial animation, for a group of objects. It is more convenient than traditional frame animation. However, the elastic object movement is not attached to skeleton animation. Furthermore, this elastic simulation is not in real time. A possible future work that can be done based on the elastic simulation is to define a skeleton system and to map the mesh body onto it. The different parts of the body can be defined as the different freedom of deformable based on the elasticity. For example, the mesh is less elastic on the arms, legs; the mesh is more elastic on the areas that consist fats, like breast, belly. The weight of the elastic property of the muscles can be mapped and dynamically set according to the skeleton. The system can be integrated into advanced animation software as a plug-in.

- Collision Detection between soft objects is a complex phenomenon, which has not been widely developed in physics. In our current system, we are using the penalty methods, which do not generate the contact surface between the interacting objects [7]. This method uses the amount of inter-penetration for computing a force which pushes the objects apart instead. Even though the result is fair enough based on estimation, in reality, the contact surfaces should be generated rather than local inter-penetrations. Especially, if we want to use computer animation to imitate organ surgery and help surgeon practice as if interact with real objects, the penalty method is no longer appropriate. There must be a more accurate algorithm to define the collision between rigid body and soft body, or soft body to soft body. Our software should be able to describe other soft body deformation, such as fractures.

- Integration with the key-frame based animation of a softbody parts on a skeleton or a Bezier curve with the body center mass attached to the curve.

- Integration with CUGL.

- Implement and compare the ENO schemes.

- Non-Uniform Unit Sphere

# Bibliography

[1] D. Baraff and A. Witkin. Dynamic simulation of non-penetrating flexible bodies. *Computer Graphics*, 26(2):303–308, 1992.

[2] N. D. and R. Lobb. A fluid-based soft-object model. *Comp. Graph. and App*, 22(4):68–75, July 2002.

[3] F. Gibson and B. Mirtich. A survey of deformable models in computer graphics. Technical Report TR97–17, Mitsubishi Electric Research Laboratories, November 1997.

[4] J. Hodgins and J. O'Brien. Computer animation. In *Wiley Encyclopedia of Electrical and Electronics Engineering*, volume 3, pages 686–690. John Wiley and Sons, 1999.

[5] M. Matyja. Inverse dynamic displacement constraints in real-time cloth and soft-body models in graphics programming methods. In *Graphics Programming Methods*. Charles River Media, Inc., 2003.

[6] M. Matyja. A pressure model for soft body simulation. In *Svenska Föreningen för Grafisk Databehandling (SIGRAD2003)*, November 2003.

[7] M. Moore and J. Wilhelms. Collision detection and response for computer animation. *Computer Graphics*, 22(4):289–298, August 1988.

[8] M. Song. Dynamic Deformation of Uniform Elastic Two-Layer Objects. Master's thesis, Department of Computer Science and Software Engineering, Concordia University, Montreal, Canada, Aug. 2007.

[9] M. Song and P. Grogono. A framework for dynamic deformation of uniform elastic two-layer 2D and 3D objects in OpenGL. In *Proceedings of C3S2E'08*, Montreal, Quebec, Canada, May 2008. ACM and BytePress. To appear, `https://confsys.concordia.ca/c3s2e-08`, ISBN 978-1-60558-101-9.

[10] D. Terzopoulos, J. Platt, and K. Fleischer. From gloop to glop: heating and melting deformable objects. In *Graphics Interface '89*, pages 219–226, 1989.

[11] F. Thomas and O. Johnston. *Disney Animation: The Illusion of Life*. Abbeville Press, 1984.

[12] Wikipedia. *Procedural Modeling*. `http://en.wikipedia.org/wiki/`, 2007.

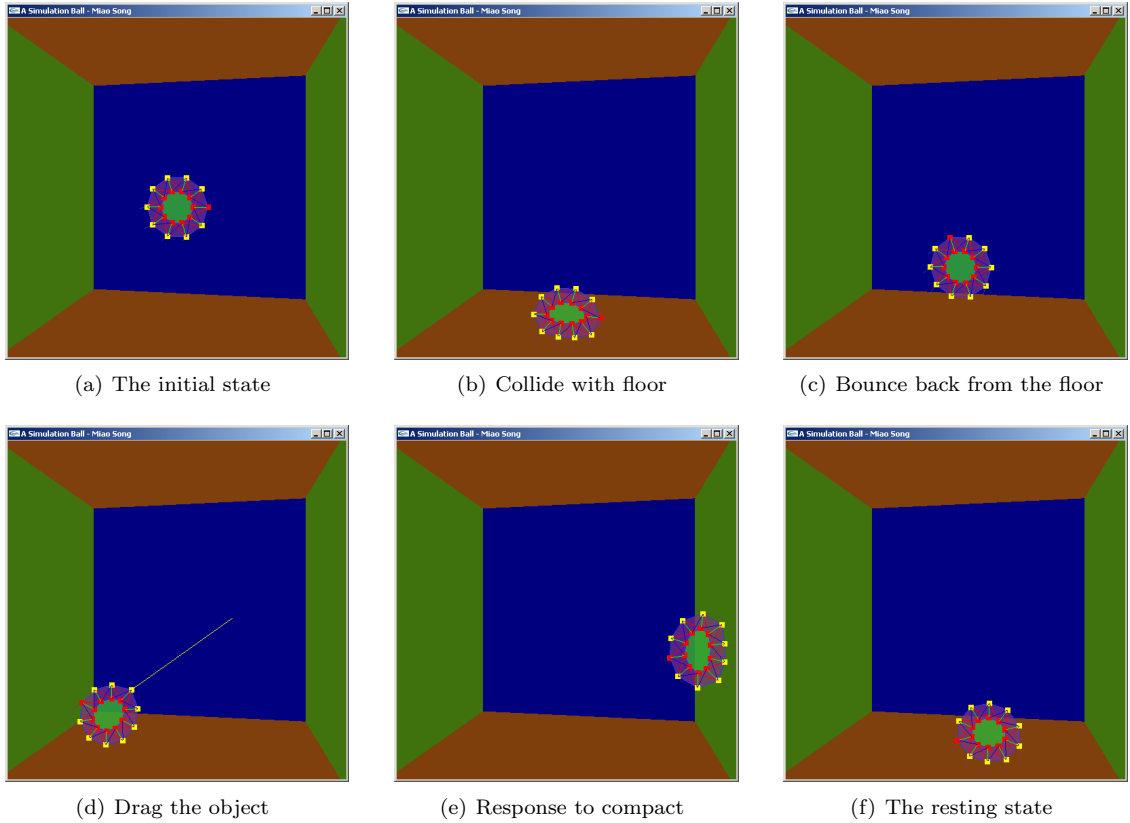

Figure 5.1: Animation Sequence of Two Dimensional Elastic Object

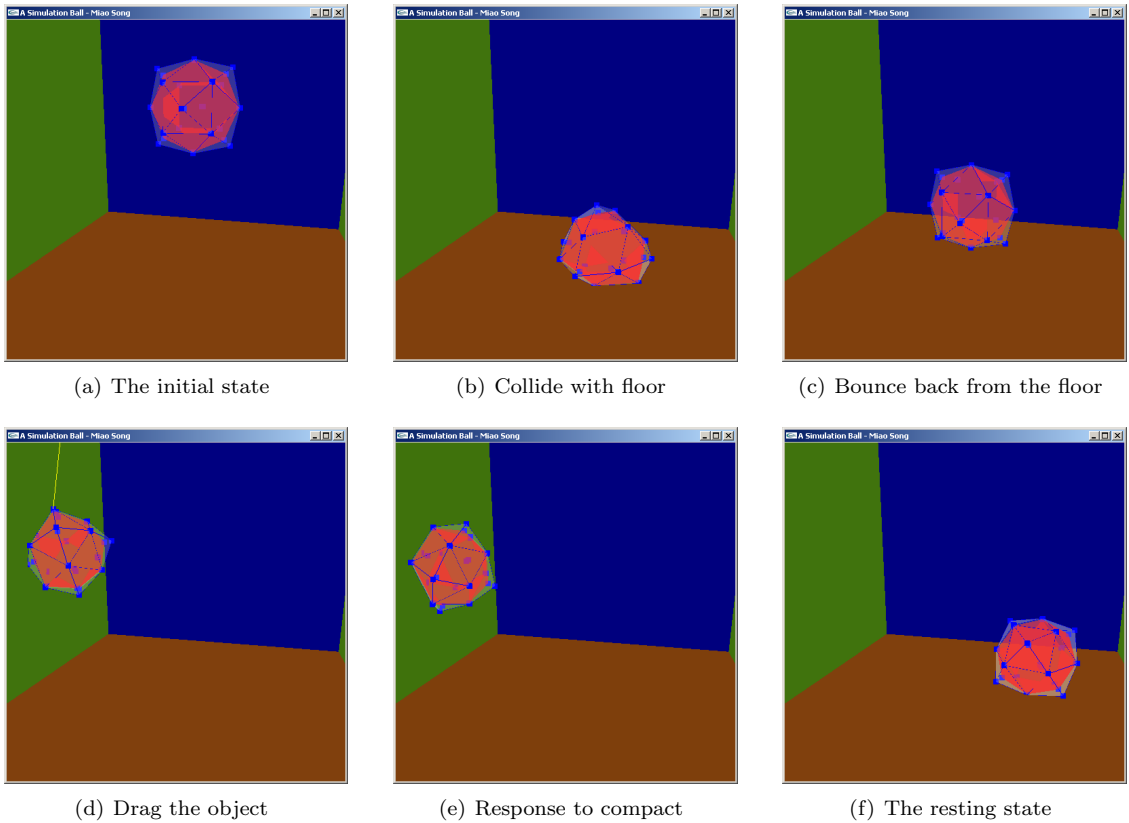

Figure 5.2: Animation Sequence of Three Dimensional Elastic Object

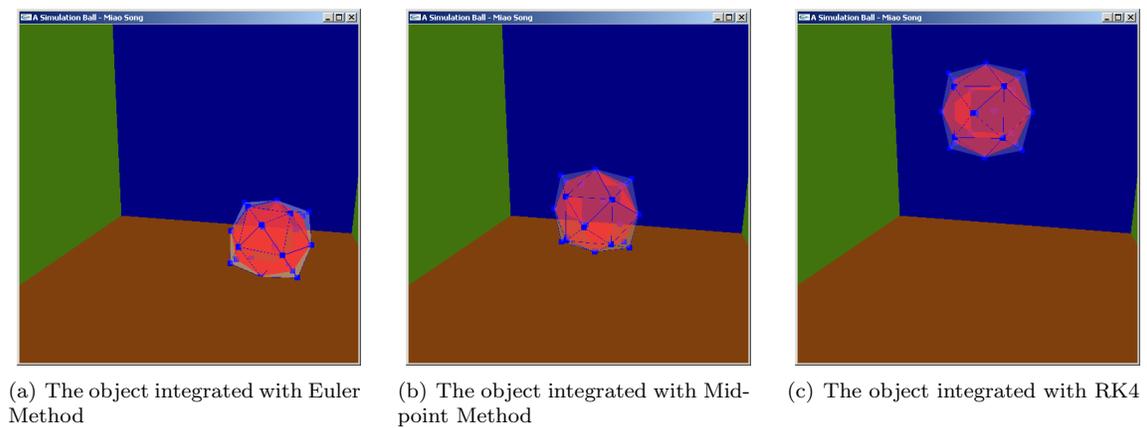

(a) The object integrated with Euler Method
(b) The object integrated with Midpoint Method
(c) The object integrated with RK4

Figure 5.3: Elastic Object at Timestep = 0.003

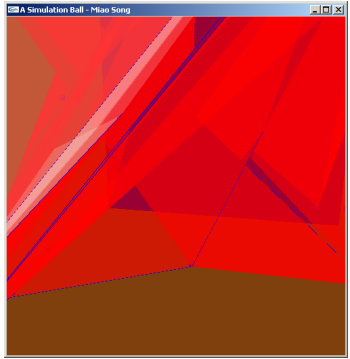
(a) The object integrated with Euler Method

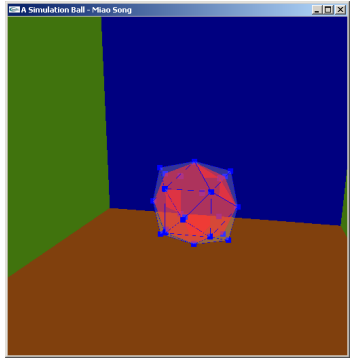
(b) The object integrated with Midpoint Method

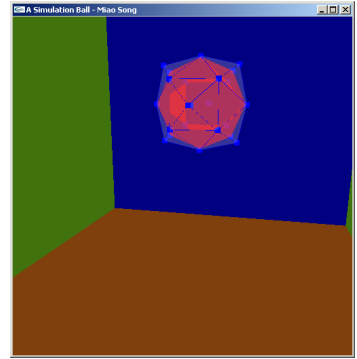
(c) The object integrated with RK4

Figure 5.4: Elastic Object at Timestep = 0.03

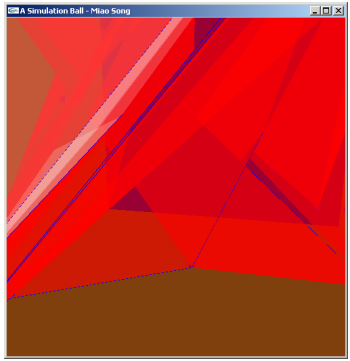
(a) The object integrated with Euler Method

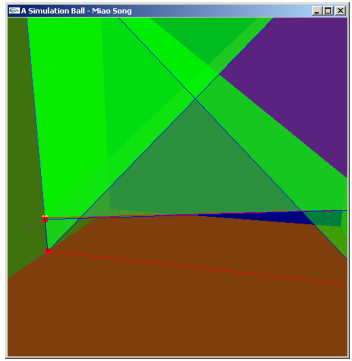
(b) The object integrated with Midpoint Method

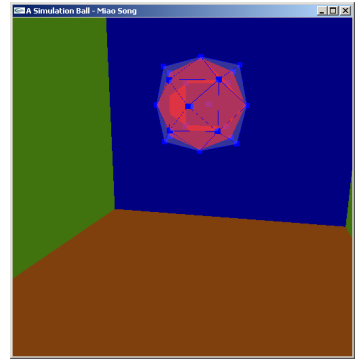
(c) The object integrated with RK4 Method

Figure 5.5: Elastic Object at Timestep = 0.3

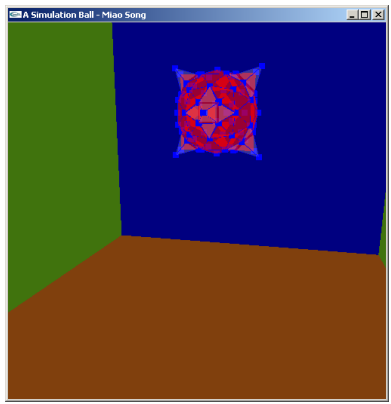

Figure 5.6: Second Subdivision Iteration

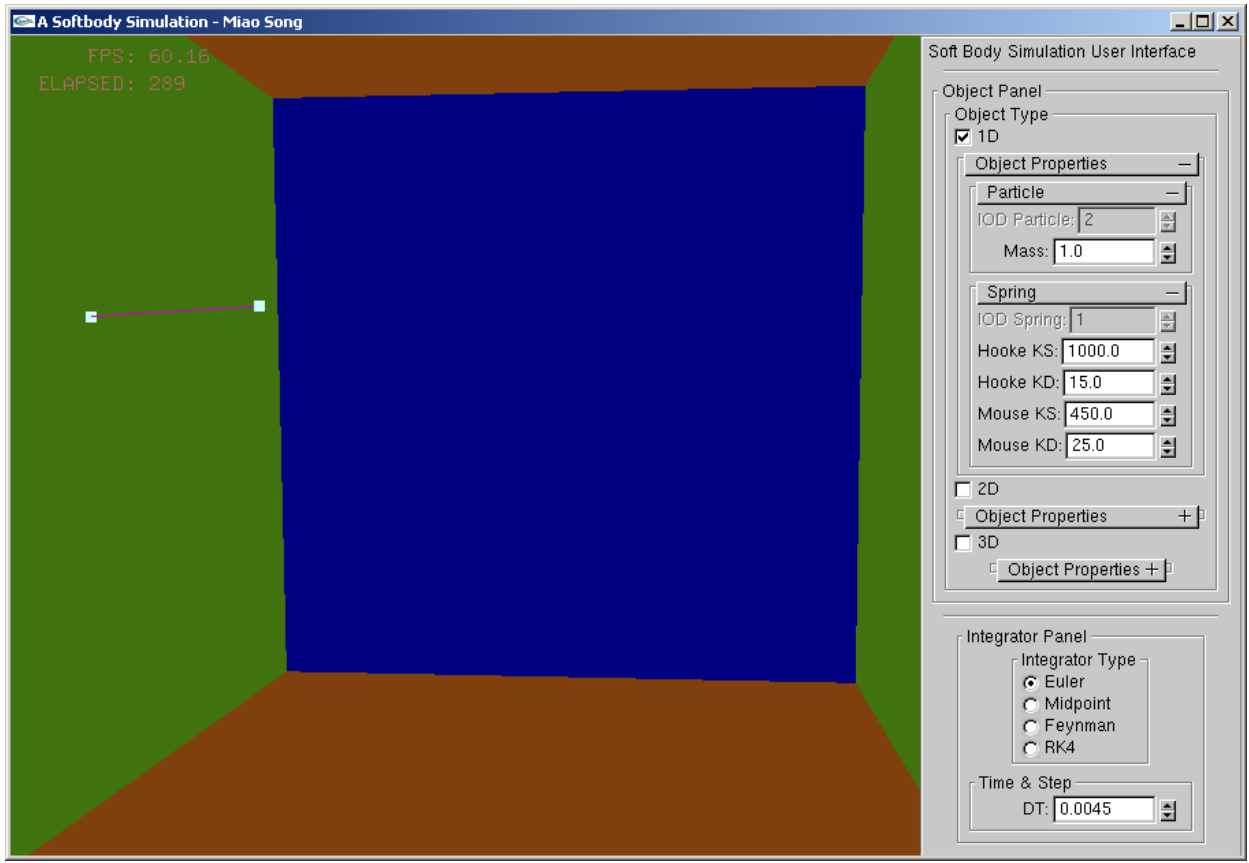

Figure 5.7: GUI for 1D Object Animation In-progress

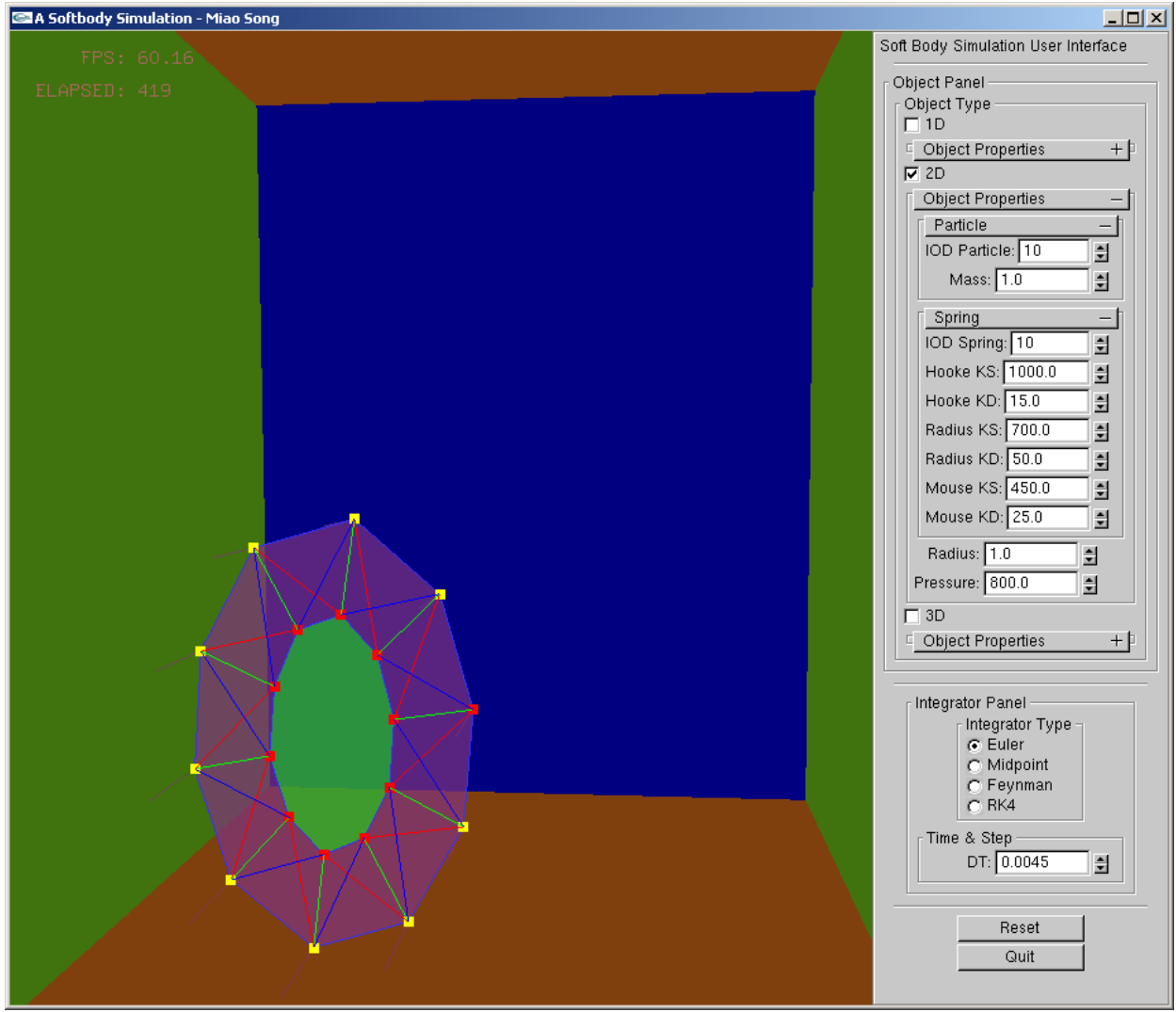

Figure 5.8: GUI for 2D Object Animation In-progress

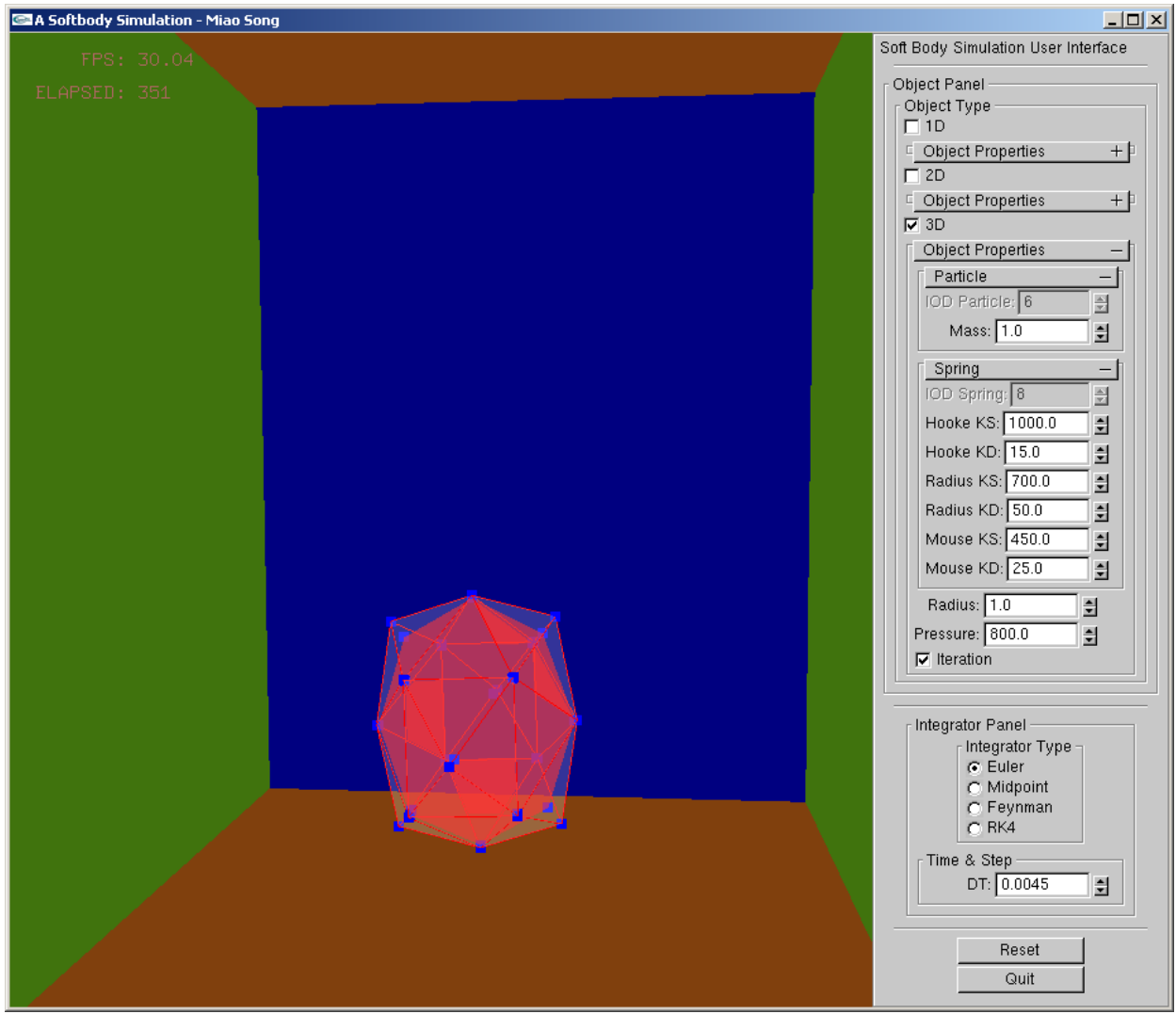

Figure 5.9: GUI for 3D Object Animation In-progress

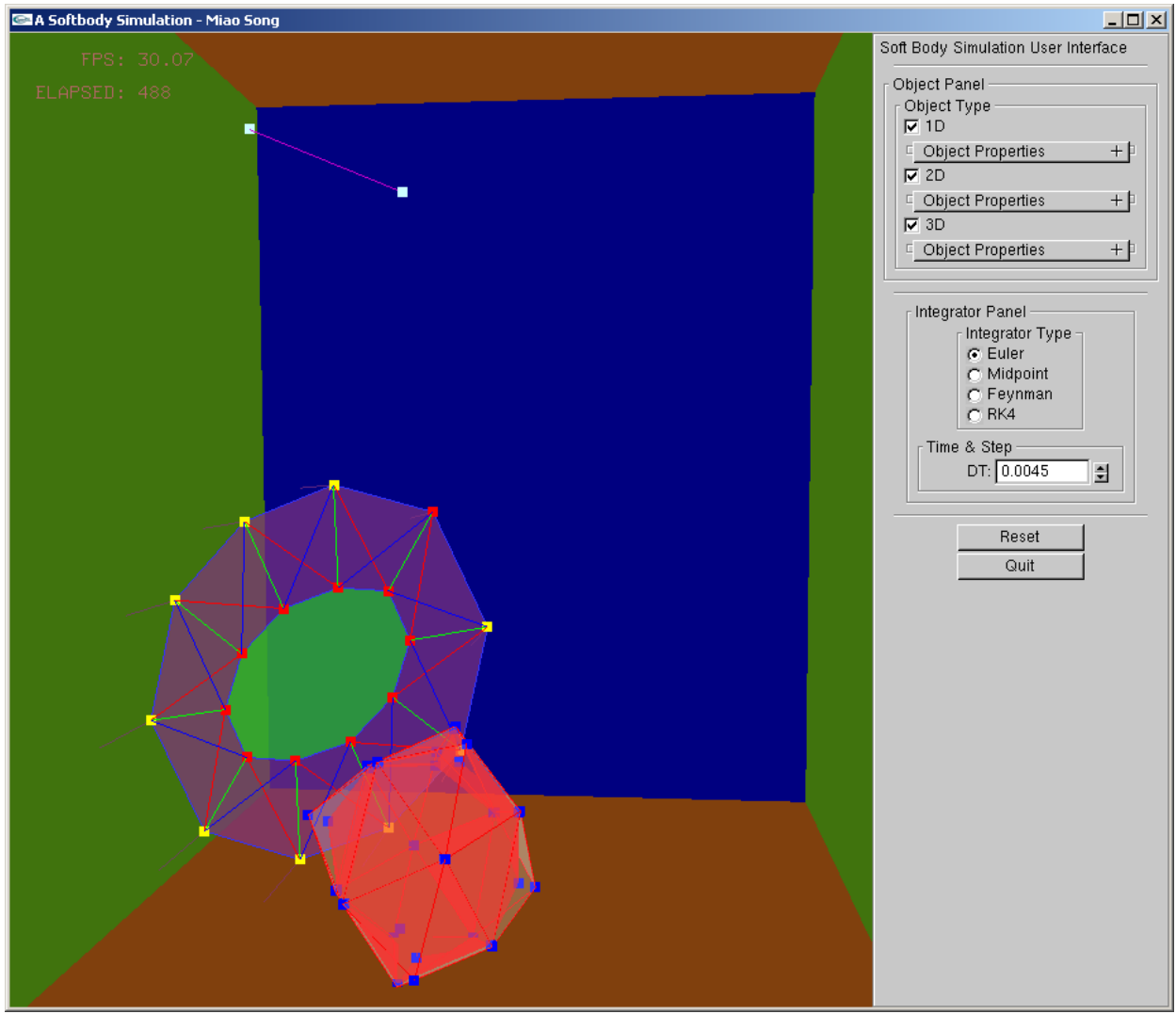

Figure 5.10: GUI for 1D, 2D, and 3D Object Animation Together In-progress